\documentclass[useAMS,usenatbib]{mn2e}
\usepackage{amssymb}
\input{psfig.sty}

\title { Neural network correction of astrometric chromaticity } 
\author[M. Gai and R. Cancelliere]{M. Gai$^{1}$ 
\thanks { E-mail: gai@to.astro.it (MG), cancelli@di.unito.it (RC)} 
and R. Cancelliere$^{2}$ \\ 
$^{1}$Istituto Nazionale di Astrofisica -- Osservatorio Astronomico di 
Torino, V. Osservatorio 20, 10025 Pino T.se (TO), Italy \\ 
$^{2}$Dipartimento di Informatica, Universit\`a di Torino, 
C.so Svizzera 185, 10149 Torino, Italy }

\begin{document}

\date{Accepted ... Received ...; in original form ... }

\pagerange{\pageref{firstpage}--\pageref{lastpage}} 
\pubyear{2005}

\maketitle

\label{firstpage}

\begin{abstract}
In this paper we deal with the problem of chromaticity, i.e. apparent 
position variation of stellar images with their spectral distribution, 
using neural networks to analyse and process astronomical images. 
The goal is to remove this relevant source of systematic error in the 
data reduction of high precision astrometric experiments, like Gaia. 
This task can be accomplished thanks to the capability of neural networks 
to solve a nonlinear approximation problem, i.e. to construct an 
hypersurface that approximates a given set of scattered data couples. 
Images are encoded associating each of them with conveniently chosen 
moments, evaluated along the $y$ axis. 
The technique proposed, in the current framework, reduces the initial 
chromaticity of few milliarcseconds to values of few microarcseconds. 
\end{abstract}


\section{Introduction}
\label{intro}

The location of the position of a stellar image is possible with accuracy 
well below its characteristic size, when the signal to noise ratio ($SNR$) 
is sufficiently high. 
The location uncertainty is $\sigma = \alpha \cdot L / SNR$, where 
$\alpha$ is a factor keeping into account geometric factors and the 
centring algorithm, and $L$ is the root mean square width of the image 
\cite{gai2}. 
The best estimate of image position is obtained by a least square 
approach, evaluating the discrepancy between the data and the template 
describing the reference image. 
The location algorithm is therefore very sensitive to any variation of 
the actual image with respect to the selected template. 

It is necessary to check the compatibility between the real image and 
the reference profile; also important is the capability of extracting 
from the data a set of parameters suitable for a new definition of the 
template, in order to improve its consistency with the data. 
Self-calibration of the data, by deduction of the parameters for 
optimisation of the image template, is a key element in the control 
of the systematic effects in the position measurement. 

In particular, the individual spectral distribution of each object 
results in a signature on the image profile, due to diffraction, above 
all in presence of aberrations. 
Because of these reasons, our target is the implementation of a tool for 
analysis of realistic images.
 
Attempts to use neural networks (NN) in astronomy have been performed in 
the past, mainly in the field of adaptive optics: details can be found 
e.g. in \cite{Loyd} and \cite{Wizin}. 

In Section \ref{imaging} we discuss the image characterisation problem 
addressed in the present work; in Section \ref{NeuralNets} we resume the 
main features of sigmoidal NN and backpropagation algorithm, with a 
brief reminder of the specific definitions, and in Section \ref{processing} 
we describe the generation of the data set, its processing and the 
current results.

\section { Diffraction imaging }
\label{imaging}

The image of a star, considered as a point-like source at infinity, 
and produced by an ideal telescope, is derived in basic textbooks 
on optics. 
For an unobstructed circular pupil of diameter $D$, at wavelength 
$\lambda$, it has radial symmetry and is described by the squared 
Airy function \ref{eq:airy} (see \cite{born} for notation). 
\begin{equation} 
\label{eq:airy} 
I\left( r\right) = k \left[
{2\,J_{1} \left( r\right) } \, / \, r \right] ^{2} \, .  
\end{equation}
Here $J_1$ is the Bessel function of the first kind, order one, 
$k$ a normalisation constant, and $r=D/2$ the aperture radius. 
The Airy diameter, between the first two minima, is $2.44 \lambda / D$ 
in angular units; the linear scale is defined by the focal length. 

The diffraction image on the focal plane of {\it any} real telescope, 
described by a set of aberration values, for a given pupil geometry, 
is deduced by the square modulus of the Fourier transform of the 
pupil function $e^{i \Phi}$: 
\begin{equation}
\label{eq:PSF}
I\left( r,\phi \right) =\frac{k}{\pi ^{2} }
\left| \int d\rho  \int d\theta  \,\rho 
\,e^{i\Phi \left( \rho ,\theta \right) } e^{-i\pi r\rho \,\cos \left(
\theta -\phi \right) } \right| ^{2} 
\end{equation}
where $\left\{r,\ \phi\right\}$ and $\left\{\rho,\ \theta\right\}$ 
are the radial coordinates, respectively on image and pupil plane, 
and the integration domain corresponds to the pupil: for the circular 
case, $0 \le \rho \le 1; \, 0 \le \theta \le 2\pi$. 
In case of a rectangular pupil, it is more convenient to use cartesian 
coordinates on both image and pupil plane, e.g. $\left\{ x, \, y \right\}$ 
and $\left\{ \xi, \, \eta \right\}$, respectively, integrating between 
the appropriate boundaries $\left[ \xi_1, \, \xi_2 \right]; \, 
\left[ \eta_1, \, \eta_2 \right]$. 
\\ 
The phase aberration $\Phi$ describes for the real case the deviation 
from the ideal flat wavefront, i.e. the wavefront error (WFE), and is 
usually decomposed by means of a set of functions (e.g. the five 
Seidel classical aberrations or Zernike functions, whose first 21 terms 
are listed in Tab. \ref{tab_zer}): 
\begin{equation}
\label{eq:aberr}                                                                                       
\Phi \left( \rho ,\theta \right) = \frac{2\pi}{\lambda} WFE = 
\frac{2\pi}{\lambda} \sum_{n=1}^{21} A_n \phi_n (\rho,\theta) \ . 
\end{equation}
If $\Phi = 0$ (non-aberrated case, $\{A_{n}\} = 0$), we obtain a flat 
wavefront, i.e. $WFE = 0$, and Eq. (\ref{eq:airy}) is retrieved for 
the circular pupil. 

The nonlinear relation between the set of aberration coefficients $A_n$ 
and the image is put in evidence by replacement of Eq. (\ref{eq:aberr}) 
in Eq. (\ref{eq:PSF}). 
In particular, the WFE is independent from wavelength, and wavelength 
dependence in the pupil function is shown by the $2 \pi / \lambda$ 
factor. 
\\ 
The real polychromatic image of an unresolved stellar source is produced 
by integration over the appropriate bandwidth of the monochromatic PSF 
above, weighed by the combination of source spectral distribution, 
instrument transmission and detector response. 
Thus, objects with different spectral distributions have different 
image profiles, and the position estimate produced by any location 
algorithm (e.g. the centre of gravity, COG, or barycentre) is affected 
by discrepancy with respect to the nominal position from the image 
generated by an ideal optical system. 

The variation of apparent position with source spectral distribution is 
what we call {\it chromaticity}, and it is relevant to high precision 
astrometry because in normal telescope configurations it can amount to 
several milliarcseconds, inducing severe limitations with respect to the 
measurement goal. 
For example, in the Gaia mission \cite{Gaia}, the individual exposure 
precision for bright objects is of order of few ten microarcseconds. 
It is possible to use different position estimators (e.g. least square 
methods rather than COG), and each procedure is affected by a specific 
spectral sensitivity. 

\begin{table}
\begin{center}
\begin{tabular}{llll}
  \hline 

1 & 1 & 12 & $(4\rho^{4}-3\rho^{2})cos(2\theta)$ \\ 
2 & $\rho cos(\theta)$ & 13 & $(6\rho^{4}-6\rho^{2}+1)$ \\ 
3 & $\rho sin(\theta)$ & 14 & $(4\rho^{4}-3\rho^{2})sin(2\theta)$ \\ 
4 & $\rho^{2}cos(2\theta)$ & 15 & $\rho^{4}sin(4\theta)$ \\
5 & $2\rho^{2}-1$ & 16 & $\rho^{5}cos(5\theta)$ \\ 
6 & $\rho^{2}sin(2\theta)$ & 17 & $(5\rho^{5}- 4\rho^{3})cos(3\theta)$ \\ 
7 & $\rho^{3}cos(3\theta)$ & 18 & $(10\rho^{5}-12\rho^{3}+3\rho)cos(\theta)$ \\
8 & $(3\rho^{3}-2\rho)cos(\theta)$ & 19 & $(10\rho^{5} -  12 \rho^{3} + 3 \rho) sin(\theta)$ \\ 
9 & $(3\rho^{3}-2\rho)sin(\theta)$ & 20 & $(5\rho^{5}-4\rho^{3})sin(3\theta)$ \\

10 & $\rho^{3}sin(3\theta)$ & 21 & $5\rho^{5}sin(5\theta)$ \\
11 &  $\rho^{4}cos(4\theta)$ & & \\ 
  \hline 
\end{tabular}
\caption {The 21 lowest order Zernike polynomials }
\label{tab_zer}
\end{center}
\end{table}

The common misconception that reflective optics is ``achromatic'' is 
true in the sense that it is not affected by classical chromatic 
aberration, typical of refractive systems. 
However, chromaticity in the above sense is critical. 
Also, not all aberrations are relevant to chromaticity, but the 
relationship is not mathematically trivial; the critical terms 
introduce an asymmetry in the image, along the measurement direction, 
and are associated to odd parity functions. 
An analysis of chromaticity versus aberrations, optical design aspects, 
and optical engineering issues, are discussed in a separate paper, in 
preparation, which also deals with design optimisation guidelines. 
After minimisation of the chromaticity by design and construction, the 
residual chromaticity must be taken into account in the data reduction 
phase. 

The aberration components are not easily measured during operation. 
In principle, it is possible to use techniques developed in past works 
\cite{canc2} for aberration reconstruction from the focal plane images. 
This may be considered for future work, but given the number of 
aberrations terms and quickly increasing size of the data set of examples 
required for proper training, the computational load becomes quite large. 

Instead, in the current paper we are interested in the classification 
capability of a NN to implement identification of the chromatic effect 
from the image profile itself, and subsequent correction in the data 
reduction. 
The goal is a tool for chromaticity self-calibration throughout the 
mission, crucial with respect to high precision astrometry. 
We find that the image moments are convenient description parameters, 
as discussed below. 

The chromaticity is estimated as difference between the COG of a blue (B3V) 
and red (M8V) stars, modelled as black bodies, with effective wavelengths 
628~nm and 756~nm respectively, deduced by taking into account also the 
telescope transmission and detector quantum efficiency. 
A set of aberration cases is generated for the basic telescope geometry of 
Gaia (i.e. $0.49~m$ off-axis, $1.4 \times 0.5~m$ aperture), under the 
assumption of small image degradation, i.e. of reasonably good imaging 
performance, as desired for large field astronomical telescopes. 
The aberration coefficients are generated with a uniform random distribution 
with peak value 50~nm for each component, using the Zernike formulation. 
The coefficient range is not specific of a given configuration, but 
represents all mathematically possible cases, i.e. a superset of the 
optically feasible systems.

\subsection {Image encoding} 

To maximise the field of view, i.e. observe simultaneously a large 
area, typical astronomical images are sampled over a small number 
of pixels. 

The minimum sampling requirements, related to the Nyquist-Shannon 
criterion, are of order of two pixels over the full width at half 
maximum, or about five pixels within the central diffraction peak. 
The signal detected in each pixel is then affected by strong 
variations 
depending on the initial phase (or relative position) of the 
parent 
intensity distribution (the continuous image) with respect to the 
pixel array, even in a noiseless case. 
The pixel intensity distribution of the measured images, then, 
is not convenient for evaluating the discrepancy of the effective 
image with respect to the nominal image. 

It may be possible to add a sort of magnifying device, providing good 
sampling for the images in a small region: in this case, the resolution 
is adequate to minimise the effects of the finite pixel size \cite{gai}. 
In Gaia this would have an heavy impact on the payload, so that we focus 
on methods applicable directly to the science data. 
Even in case of well sampled images, we have to face some problems: 
assuming a sampling of 20 pixels per Airy diameter, and reading up to 
the third diffraction lobe, the image size is $60 \times 60 = 3600$ 
pixels. 
Direct usage of such images as input data to the NN is impractical, 
because of the large computational load involved, and identification 
of a more compact encoding, using the science data rather than 
additional custom hardware, appears to be appealing. 

Since the Gaia measurement is one-dimensional, and most images are 
integrated in the across scan direction, the problem (and the signals 
considered) is also reduced to one dimension, conventionally labelled 
$y$: the one-dimensional image is $ I\left( y\right)$. 
The encoding scheme we adopt for the images allows extraction of the 
desired information for classification; each input image is described by 
the centre of gravity and the first central moments as follow:
 
\begin{equation}
\begin{array}{l}
\mu _{y} = \int dy\,y\cdot I\left( y\right)  
\ \left/ \right. 
I_{int} 
\\[3mm]
\sigma _{y}^{2} = \int dy\,\left( y-\mu _{y} \right) ^{2} \cdot
I\left( y\right)  
\ \left/ \right. 
I_{int} 
\\[3mm] 
M(j) = \int dy\, \left( \frac{y-\mu _{y} }{\sigma _{y} }
\right) ^{j} \cdot I\left( y\right) 
\ \left/ \right. 
I_{int}, \ \ \  j > 2 
\end{array} 
\label{eq:moments}
\end{equation}
\noindent 
where $I_{int} = \int dy\,I\left( y\right)$ is the integrated photometric 
level of the measurement. 

One-dimensional encoding is a further change with respect to previous 
problems, in which we took advantage of the full two-dimensional image 
structure to deduce the different aberration terms. 
\\ 
The central moments are much less sensitive than the pixel intensity 
values to the effects related to the finite pixel size and the 
position of the image peak with respect to the pixel borders, i.e. 
the relative phase between optical image and pixel array. 
Thus, central moments can be deduced conveniently also on the detected 
low resolution images, without the need for high resolution detectors. 
The encoding technique based on using moments as image description 
parameters for neural processing was first introduced in \cite{canc}, 
where more details are available.

\section { Sigmoidal Neural Networks }
\label{NeuralNets}

Neural networks learn from examples that is, 
given the training set of $N$ multi-dimensional data pairs 
$\{\left( x_i, F\left(  x_i \right)\right) /  x_i \in \mathbb R^P, F\left( x_i 
\right) \in \mathbb R^Q \}, \ i=1, \dots, N, \ $ after the training if $x_i$ is 
the input to the network, the output is close to, or 
coincident with, the desired answer $F\left(  x_i \right)$ and the network has 
generalization properties too, that is it gives 
as output $F\left( x_i \right)$ even if the input is only ``close to''  
$x_i$, for instance a noisy or distorted or incomplete version of $x_i$; 
a comprehensive review on NN properties and applications can be found in 
\cite{hay}. 

In our work we use the multilayer perceptron, first introduced in 1986 (see \cite{rum}), as an 
extension of the perceptron model \cite{min}.

\begin{figure}
\centerline{\psfig{figure=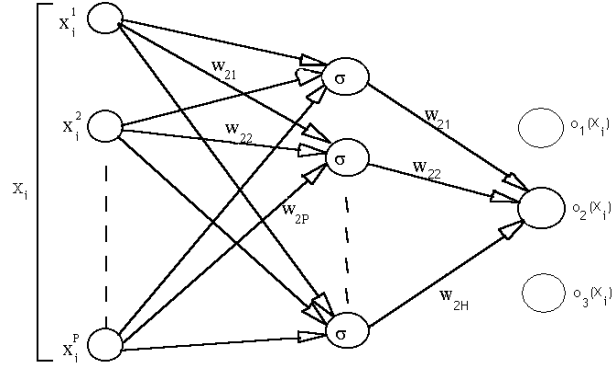,height=6cm}}
\caption{A multilayer perceptron with one hidden layer}
\label{figura1}
\end{figure}

The multilayer perceptron, with sigmoidal units in the hidden layers, 
is one of the most known and used NN model: it computes distances in 
the input space (i.e. among patterns $x_i \in \mathbb R^P$) using a 
metric based on inner products and it is usually trained by the 
backpropagation algorithm.
The architecture of a sigmoidal NN is schematically shown in 
Fig. \ref{figura1}, in which we find the most common three-layers case. 
The network is described by Eqs. \ref{eq:network}: 
\begin{equation}
\begin{array}{l}
a^{k+1}_{j} = \sum\limits_{j'}w_{jj'}o^k_{j'}+bias_j
\\[4mm]
o^{k+1} _{j} = \sigma (a^{k+1}_{j}) \equiv \frac{1}{1+e^{-a^{k+1}_{j}}}
\\[4mm] 
o^{out} _{m} \equiv \sum\limits_{j}w_{mj} o^{out-1}_{j} 
\end{array} 
\label{eq:network}
\end{equation}

Here $a$ is the input to each unit, $o$ is its output and $w_{ij}$ is 
the weight associated to the connection between units $i$ and $j;$ 
each unit is defined by two indexes, a superscript specifying 
its layer (i.e. input, hidden or output layer) and a subscript 
labelling each unit in a layer. 
\\ 
The training procedure is finalized to find the best set of weights 
$\{w_{ij}\}$ 
solving the approximation problem $o\left(  x_i \right) \approx F\left( 
x_i \right) $ and this is usually reached by the  iterative process 
corresponding to the standard backpropagation algorithm. 

At each step, each weight is modified accordingly to the gradient descent 
rule (a more detailed description can be found in \cite{rum}), completed 
with the momentum term, 
$w_{ij} = w_{ij} +\Delta w_{ij} $, 
$\Delta w_{ij} = - \eta \frac{\partial E}{\partial w_{ij} }$ 
where $E$ is the error functional defined above.\\ 
This procedure is iterated many times over the complete set of 
examples $\{x_i, F\left(  x_i \right)\}$ (the training set), 
and under appropriate conditions it converges to a suitable set of 
weights defining the desired approximating function. 
Convergence is usually defined in terms of the error functional, 
evaluated over the whole training set; when a pre-selected 
threshold $E_T$ is reached, the NN can be tested using a different 
set of data $\{x_i', F\left(  x_i' \right)\}$, the so called test set.

\section{Data processing and results}
\label{processing}

\begin{figure}
\centerline{\psfig{figure=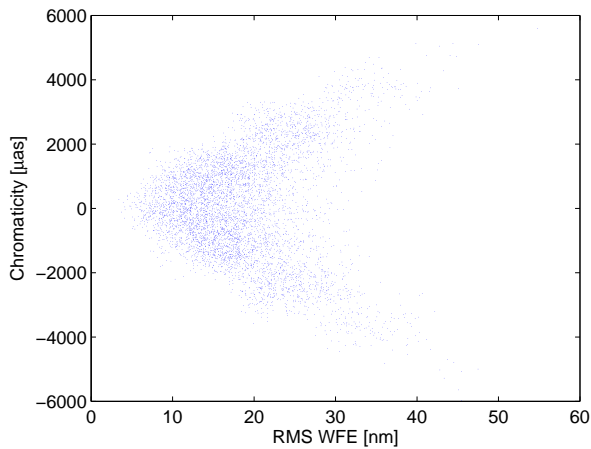,height=6cm}}
\caption { Distribution of chromaticity vs. RMS WFE over the test set. }
\label{figwfe}
\end{figure}

In this section we describe the identification of the most convenient 
image parameters, the generation of the training and test sets, and the 
results from the NN processing. 
The sources are represented by the monochromatic PSF at the wavelength of 
respectively 628~nm (B3V) and 756~nm (M8V), deduced from the blackbody 
spectrum associated to the effective temperature of each star, and the 
expected spectral distributions of instrument transmission and quantum 
efficiency.

\subsection{Aberration sample} 

In order to investigate the relationship among image moments and 
chromaticity, we start from a reasonable sampling of the aberration 
space, using a uniform random distribution of the 21 lowest order 
Zernike coefficients (Tab \ref{tab_zer}), within the range $\pm 50$~nm 
on each term. 
\\ 
For each aberration case, defined by the set of 21 Zernike coefficient 
values, we evaluate the RMS WFE for verification purposes and we build 
the PSF for the two source cases; on the PSF, the photo-centre position 
is evaluated as the COG, and the moments up to order five are computed 
accordingly to the definitions in Eq. (\ref{eq:moments}), after across 
scan integration to replicate the Gaia measurement process. 
The chromaticity is directly derived as COG difference. 
\\ 
In Fig.\ref{figwfe} we show the distribution of chromaticity vs. RMS WFE 
over the test set (5821 instances). 
At increasing values of the aberration RMS WFE, the chromaticity has 
usually larger absolute value, but the relationship is not simple; the 
same consideration holds for the relationship between chromaticity and 
any other image moment, due to diffraction non-linearity. 
Some statistical parameters of the distribution of chromaticity and WFE 
values in the training data set are listed in Tab. \ref{tab_wfe}. 

\begin{figure}
\centerline{\psfig{figure=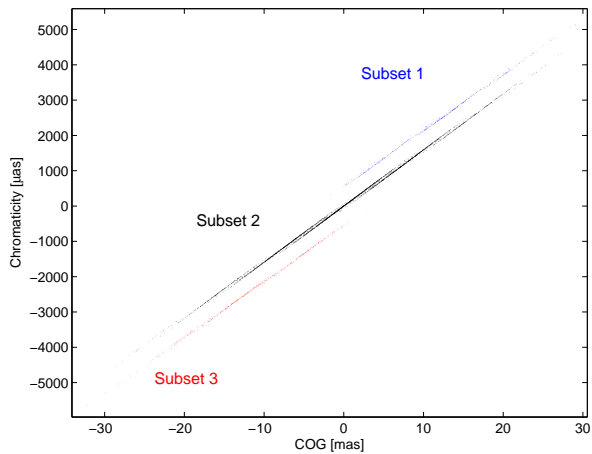,height=6cm}}
\caption { Distribution of chromaticity vs. image COG over the test set. }
\label{figCOG}
\end{figure}

The average RMS WFE corresponds to an overall optical quality of about 
$\lambda$/30 at 600~nm, i.e. a comparably good performance; some of the 
optical designs considered for Gaia provide a RMS WFE of about 40~nm, or 
$ \lambda$/15 at 600~nm. 
The chromaticity evaluated on the proposed designs has peak values of 
$\sim 2-3$~mas, localised in specific field positions, and symmetric 
distribution for the nominal aligned configuration. 
The random data set considered is thus reasonably representative of 
a range of realistic optical configurations. 

\begin{table} 
\begin{center}       
\begin{tabular}{ccc} 
\hline
	& Chromaticity 	& RMS WFE 	\\ 
	& [$\mu$as] 	& [nm] 	\\ 
\hline
Min. 	& -5289.3 		& 2.99 	\\ 
\hline
Mean 	& 6.1 		& 18.71 	\\ 
\hline
Max.	& 5365.9 		& 46.26  	\\ 
\hline
RMS	& 1648.2 		& 6.81  	\\ 
\hline
\end{tabular}
\caption { Statistics over the training data set of chromaticity 
and RMS WFE. }
\label{tab_wfe} 
\end{center}
\end{table}

\subsection{Neural network input}

	We verify that the across scan moments ($x$ in the Gaia reference frame) 
are all irrelevant, i.e. their effect on chromaticity is negligible. 
Usage of the standard one-dimensional science data is therefore appropriate, 
without operation changes. 
The moments are all computed with straightforward operations from the 
measured data, as well as the variation with respect to the nominal 
moment values of a selected reference spectral type. 
\\ 
Some of the along scan ($y$) moments do not show an apparent signature 
associated to chromaticity. 
A few of them are still required to provide an acceptable description 
of the image profile: the moment selection was verified on the NN, 
removing some of them until reaching the minimum number of 
parameters compatible with good convergence of the training. 

\begin{figure}
\centerline{\psfig{figure=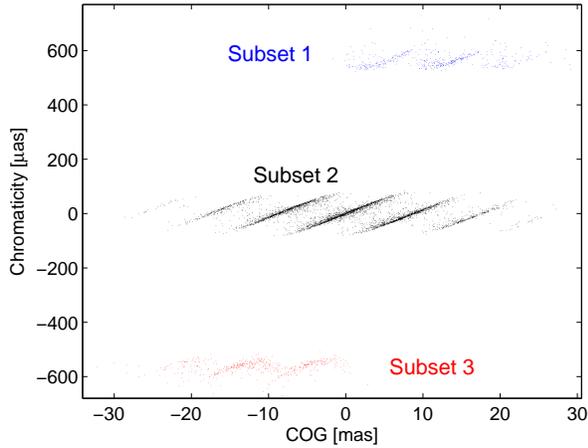,height=6cm}}
\caption { Distribution of residual chromaticity vs. image COG, after slope 
subtraction. }
\label{chresvsCOG}
\end{figure}

From the data distribution, it appears that some pre-processing is 
recommended, in order to ease the subsequent neural processing. 
This is most apparent in the distribution of chromaticity with respect 
to the nominal image COG, shown in Fig.\ref{figCOG} for the test set. 
The data points are distributed in three well-defined regions following 
parallel straight lines, shown in figure by different colours. 
\\ 
The chromaticity / COG structure is shown with even more clarity by 
subtracting the average slope, derived by linear fit on the central 
peak of the distribution. 
The fit parameters are: 157.83~$\mu as/mas$ (slope); 0.06~$\mu as$ 
(offset). 
In Fig.\ref{chresvsCOG} we show the distribution of the chromaticity 
residuals after subtraction of the above straight line. 
The number of data instances in each side peak is about 9\% of the sample. 
The peaks are quite symmetric and corresponding to $\pm 600 \ \mu$as. 
From the residuals, a finer structure appears, which is not currently 
used in pre-processing.

\begin{figure}
\centerline{\psfig{figure=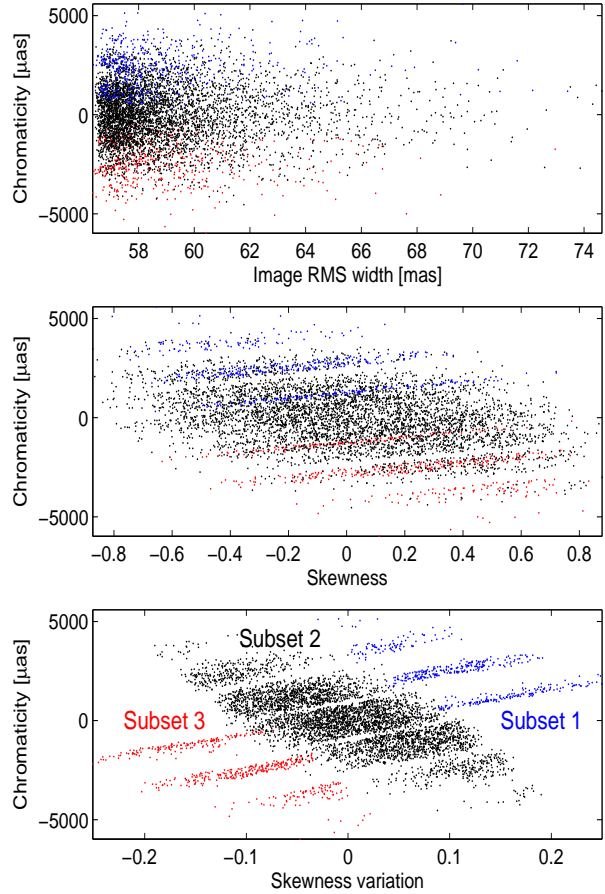,width=8cm,height=12cm}}
\caption { Distribution of chromaticity vs. image RMS width (top), skewness 
(centre) and skewness variation (bottom). }
\label{ch_mom2_3_d3}
\end{figure}

The classification of data instances in the chromaticity/COG groups 
(subsets 1, 2 and 3) is taken into account in evaluating the distribution 
of chromaticity vs. other moments. 
In some of the plots, the groups are clearly localised in specific 
parameter regions. Besides, the structure is more complex. 

Taking advantage of the structure identified on the COG distribution, 
we show in Fig. \ref{ch_mom2_3_d3} the distribution 
of chromaticity vs. image RMS width (top panel), skewness (central 
panel), and skewness variation (bottom panel) between the selected blue 
and red stars. 
The subsets are shown here with the same colours as in Fig. \ref{figCOG} 
and \ref{chresvsCOG}, i.e. blue for subset 1, black for subset 2 and 
red for subset 3. 
\\ 
The RMS width (top panel) and other even order moments do not show an 
apparent structure, and most of them are not used in the neural 
processing. 
\\ 
Odd order moments do not evidence directly the subset structure, as 
for the skewness, in the central panel of Fig. \ref{ch_mom2_3_d3}. 
Besides, the distribution of chromaticity vs. skewness variation with 
spectral type (bottom panel) clearly shows clustering of the three 
subsets. 
This appears a convenient choice for NN input, as it carries significant 
information. 
Similar effects are shown by other odd order moments. 

\begin{figure}
\centerline{\psfig{figure= 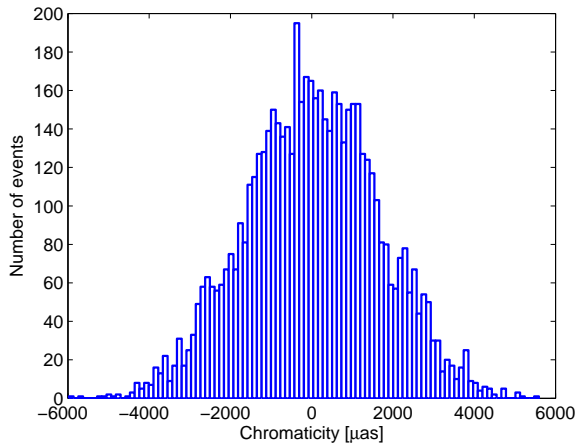,height=6cm }} 
\caption {Input chromaticity distribution. } 
\label{in_ch}
\end{figure}

The NN input can therefore be defined in terms of the local instrument 
response, encoded in the nominal moments for a reference star, and the 
individual measurement moments. 
The COG of the reference object is the deviation of the image position 
with respect to an ideal system, and it is associated to the classical 
distortion. 
The other reference object inputs are the image RMS width, the third and 
fifth order moments. 
The inputs associated to the measured signal, from an unknown type star, 
is a simple pair of values, i.e. the variation in the third and fifth 
order moments with respect to the known reference case. 
\\ 
Also, taking advantage of the data structure discussed above, we subtract 
the linear trend to the target (the chromaticity) in the training set. 
This pre-processing is supposed to ease the NN computational load. 
The inverse transformation is applied to the output data on the test set. 
\\ 
The training and test sets include respectively the data of 20000 
and 5820 aberration instances, built accordingly to the above process. 
The histogram of input chromaticity distribution in the test set 
(Fig. \ref{in_ch}) is approximately Gaussian. 

\begin{figure}
\centerline{\psfig{figure= 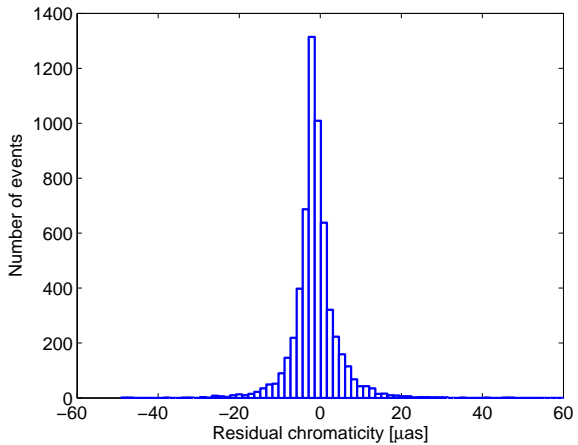,height=6cm }} 
\caption {Residual chromaticity distribution.} 
\label{result2}
\end{figure}

\subsection{Neural processing} 

We use a sigmoidal NN with six inputs (four nominal and two measured 
values), one output (the chromaticity), and a single hidden layer with 
300 units. 
The NN is optimised on the training set, and its performance is verified 
on the test set, as described in Sect. \ref {NeuralNets}. 

\begin{table} 
\begin{center}       
\begin{tabular}{lcc} 
\hline
Chromaticity & Input  		& Residual	\\ 
\hline
Min. 	[$\mu$as] & -5975.4 		& -49.1 \\ 
\hline
Mean 	[$\mu$as] & 19.0  		& -1.3 \\ 
\hline
Max.	[$\mu$as] & 5590.8 		& 100.1 \\ 
\hline
RMS	[$\mu$as] & 1641.8 		& 5.7 \\ 
\hline
Fraction in $\pm 3 \sigma [\%]$ & 99.8 & 98.0 \\ 
\hline
\end{tabular}
\caption { Statistics over the test set of input and residual chromaticity, 
and fraction of instances within $\pm 3 \sigma$. }
\label{tab_ch_in_out} 
\end{center}
\end{table}

We use an incremental training, i.e. we split the training set in four 
subsets of 5000 examples. 
In the first training phase, the NN is trained by 1000 iterations on the 
first subset, then we add the second data subset for additional 1000 
iterations on the new compound set of 10000 examples, and so on until 
including the whole training set. 
The NN training on the complete data set is carried on for a total of 
8000 iterations, with monotonic decrease of the internal overall RMS 
error on the training set. 

The NN performance is evaluated on the test set; in particular, the 
discrepancy between the NN output (estimated chromaticity) and target 
(actual chromaticity for the test set data instances) can be considered 
as the {\it residual chromaticity} after correction based on the NN 
results. 
The residual chromaticity distribution (Fig. \ref{result2}) is quite 
symmetric vs. zero, and the main statistical parameters are listed 
in Tab. \ref{tab_ch_in_out}, compared with the corresponding values 
in the input test set. 
We remark that 98\% of the output data are within the $\pm 3 \sigma$ 
interval, vs. a corresponding fraction of 99.8\% on the input. 

\begin{figure}
\centerline{\psfig{figure=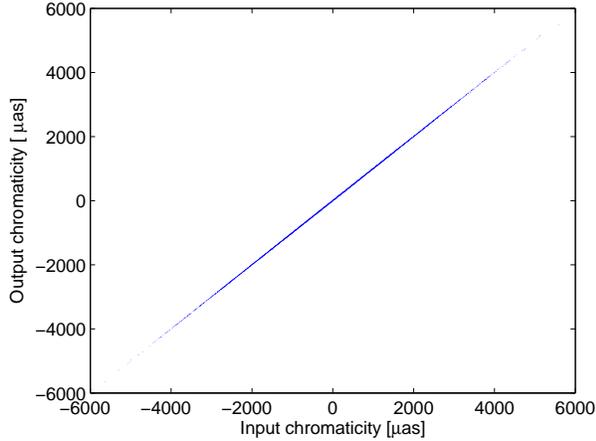,height=6cm }} 
\caption {NN performance: input/output characteristics. } 
\label{result1}
\end{figure}

Since the goal is the computation of output values coincident with the 
pre-defined target values, the characteristics, i.e. the relationship 
between input and output (plot shown in Fig. \ref{result1}) is ideally 
a straight line ($y = a + bx$) at angle $\pi/4$, passing for the origin, 
i.e. with parameters $\{a = 0, \ b = 1\}$. 
We compute the best fit parameters of the NN output vs. target 
distribution and their standard deviation; the results, shown in 
Tab. \ref{tab_stat}, are quite consistent with the expectations. 

\begin{table} 
\begin{center}       
\begin{tabular}{lc} 
\hline
Offset  & $a = 1.30 \pm 0.07$ \\ 
Slope   & $b = 1.0002 \pm 0.5e-4$ \\ 
\hline
\end{tabular} 
\caption  { Linear fit of NN output vs. input chromaticity. } 
\label{tab_stat}
\end{center}
\end{table}

\section{Conclusions}
\label{Conclusions}
In this paper we use a neural network to diagnose and correct the 
chromaticity on astrometric measurements, in a framework consistent 
with the mission Gaia. 
The science data are efficiently encoded in a set of low order image 
moments. 
The NN, with 300 internal nodes, is trained on a set of 20,000 
data instances, and evaluated on a test set of 5820 cases. 

The NN diagnostics on the test set appears to be quite effective, as 
the RMS residual chromaticity, after correction based on NN results, 
is reduced by more than two orders of magnitude (factor $\sim 280$) 
with respect to the initial RMS value (Tab. \ref{tab_ch_in_out}). 

Applying the network output for correction of the chromaticity on the 
elementary Gaia measurements, therefore, we may expect a significant 
reduction of this source of systematic error; in particular, the residual 
chromaticity can be expected to be random, and possibly subject to 
further statistical averaging in subsequent measurements. 
A word of caution is in order, however, due to the 1.3 $\mu as$ residual 
offset. 
This may not be reduced as easily by simple measurements statistics, and 
it would be desirable that this was close to zero. 
Besides, it appears that the residual offset is related to the limited 
size of the training set, and the number of internal nodes, with 
respect to the large input chromaticity range. 
Also, it may be related to the mean chromaticity of the training set, close 
to $6 \mu as$ rather than zero. 
\\ 
We expect that, increasing the training set and the number of nodes, 
the residual chromaticity offset will decrease. 
This will be part of the future developments. 
Also, the sensitivity to measurement noise, as propagated to the image 
moments, will be subject of further investigations. 

From the current results, neural network diagnostics for suppression of 
the chromatic errors on astrometric measurements appears to be a highly 
promising tool.

\end{document}